# Abrupt boundaries of intermediate phases and space filling in oxide glasses

K. Rompicharla, D. I. Novita, P. Chen, and P. Boolchand
Department of Electrical and Computer Engineering
University of Cincinnati, Cincinnati, OH 45221-0030

M. Micoulaut
Laboratoire de Physique Theorique de la Matiere Condensée, University Pierre et Marie Curie,
Boite 121, 4, Place Jussieu, 75252 Paris, Cedex05, France

W. Huff
Department of Geology, University of Cincinnati
Cincinnati, OH 45221-0013

Modulated DSC measurements on bulk $(Na_2O)_x(GeO_2)_{1-x}$ glasses show a sharp reversibility window in the 14% < x < 19% soda range, which correlates well with a broad global minimum in molar volumes. Raman and IR reflectance TO and LO mode frequencies exhibit anomalies between $x_c(1)$ = 14% (*stress* transition) and $x_c(2)$ = 19% (*rigidity* transition), with optical elasticity power-laws confirming the nature of the transitions . Birefringence measurements dramatize the macroscopically stress-free nature of the Intermediate Phase in the reversibility window.

Reversibility windows (RWs) in covalent network glasses open new avenues to understanding the structural organization and functionality of disordered matter. Glass transitions become thermally reversing and the glass forming tendency is optimized in RW. In experiments on chalcogenide alloys[1-3] changes in glass composition alter network connectivity. Experiments[1-3] supported by numerical simulations[4, 5] have shown that the sharp edges of RW define distinctive regimes of elastic behavior. Glasses in RWs form *rigid but stress-free* networks, those below the low-connectivity edge of RWs form elastically *flexible* networks, while those above the high-connectivity end of RWs form *stressed-rigid* networks. Glasses in RWs, also known as *Intermediate Phases* (IPs), have attracted much attention in recent years because they are functionally quite different from their *flexible* or *stressed-rigid* counterparts. They form space filling networks which do not age much[1-3], and are reviewed in terms of multi-scale self-organization[6] of disordered systems. A number of network systems display these characteristics of self-organization; for example, computational phase transitions[7], thin-film gate dielectrics[8], high temperature superconductors[9], H-bonded alcohols and saccharides[10] , and protein folding[11].

Is the self-organized state of disordered systems also found in ionic glasses with non-bridging oxygens as well? Alkali germanates are prototypical oxide glasses, with physical properties varying anomalously with alkali oxide content[12-14]. Here we observe a sharp RW in $(Na_2O)_x(GeO_2)_{1-x}$ glasses and show that the thermal result supported by IR reflectance and Raman scattering represents the IP. Birefringence measurements directly reveal the stress-free nature of the IP.

The mean valence number, ***r***, is a faithful measure[15] of network connectivity. In covalent systems, ***r*** is easily estimated because the valence of atoms are *locally* satisfied, leading to well defined coordination numbers[1-3]. In oxide glasses, chemical bonding has an ionic component ( $Na^+$-$O^{nb}$) in addition to a covalent one (Si-$O^b$), and the valence of ions are satisfied only on an *extended scale*[16] , making the connection between ***r*** and network connectivity more challenging. Here $O^b$ and $O^{nb}$ designate respectively bridging and non-bridging oxygen atoms.

Force fields between atoms form a hierarchy from strong nearest- neighbor forces to intermediate next-nearest –neighbor forces to weak long-range Coulombic forces. Strong forces usually serve as mechanical constraints in stabilizing a glass network[15] while weak ones usually promote only space-filling (network compaction). Coulombic forces serve to polarize a glass medium, and one observes[17] both Longitudinal- (LO) and Transverse-Optic (TO) vibrational modes, which resonate at peaks of the energy loss function Im $(-1/\varepsilon)$ (LO mode) and the imaginary part of the complex dielectric function ($\varepsilon_2$) (TO mode). Softening of a glass network upon either increasing its temperature, or upon lowering its connectivity by chemical alloying, can be elegantly probed by measuring these vibrational frequencies. We have deduced elastic thresholds and elastic power-laws in different compositional regimes of the present oxide glass system for LO and TO modes.

Fine powders of 99.995% $Na_2CO_3$ and 99.999% $GeO_2$ were weighed and intimately mixed in a nitrogen gas purged glove bag (relative humidity 5%) using a Pt crucible. Mixtures were transferred to a muffle furnace held at 125°C. The furnace T was then increased to the 1350°C-1400°C range and held there for 4 hours, and the melts were then poured onto steel plates to yield bulk glasses. Glass transitions were studied[1-3] using a TA Instrument model 2920 modulated DSC operated at 3°C/min scan rate and 1°C/100s modulation rate. Compositional trends in glass transition temperatures $T_g(x)$, non-reversing enthalpy $\Delta H_{nr}$ (x), and the specific heat jump $\Delta C_p(x)$ near $T_g$ appear in Fig.1. Densities of glasses were measured using Archimedes principle, and resulting molar volumes ($V_m$) show (Fig.1c) a broad global minimum[13, 14] near 18% (the germanate anomaly). Trends in $T_g(x)$ correlate well with a softening T of glasses when their viscosity[18] reaches $10^{11}$ Poise and not $10^{13}$ Poise, suggesting that the samples used in ref.18 probably have impurities . We observe (Fig.1b) a deep RW with abrupt edges in $\Delta H_{nr}$ (x) for 14% < x < 19%. $\Delta C_p$ (x) increases monotonically with x, but flattens in the RW to display a plateau.

Raman scattering was excited using 514 nm light and the back scattered radiation analyzed using a model 64000 Horiba Jobin-Yvon Dispersive Raman system[1-3]. By deconvoluting the line shapes of the alloyed glasses we tracked the shift in LO mode frequency with glass composition x . A Thermo-Nicolet FTIR model 900 with a Seagull accessory recorded specular reflectance over a wide range, 400 $cm^{-1}$ to 3600 $cm^{-1}$ (Fig.2). By Kramers-Kronig transformation we obtained the LO (-Im (1/ ε) ) and TO ($\varepsilon_2$) response. In both IR and Raman scattering the LO mode red-shifts with increasing x (Fig. 3a), displaying thresholds near $x_c(1)$ = 14% and $x_c(2)$ = 19%, defining three clear *regimes* of elastic behavior, for x < 14% (I), in the RW (II), and for x > 19% (III). The Raman elastic power-laws in regions I and II were deduced from

$$\nu^2 - \nu_c(1)^2 = A\,(x - x_c(1))^p \qquad (1)$$

where $\nu_c(1)$ represents the frequency at the threshold composition x = $x_c(1)$. Raman LO mode frequency ($\nu_{LO}$(Raman)) results place the power-law p in region I at $p_1$ = 1.68 (5), and in region II at $p_2$ = 0.90(5). In Fig. 3, the mode frequency $\nu_{LO}$(IR) plateaus in region II, and $\nu_{TO}$(IR) decreases in region I, jumps discontinuously near the first threshold, $x_c(1)$ = 14%, to plateau in region II, and then decreases abruptly again near the second threshold $x_c(2)$ = 19% to continue decreasing in region III along the extrapolation of region I.

Our interpretation of the results is as follows. The calorimetric RW (region II) in the present oxide glasses, as in chalcogenide glasses[1-3], also represents the IP. That view is independently confirmed by Raman optical elasticity power-laws as discussed next. In region I, $p_1 = 1.68(5)$, the same as found in sodium silicate glasses[19], compared to a value of 1.50 predicted by numerical simulations on depleted amorphous Si networks[20] for the stressed-rigid elastic phase. Region I, thus represents the stressed-rigid elastic phase of the present glasses. In region II, the Raman scattering derived LO mode power-law $p_2$ compares well with exponents of IPs measured[1-3] in chalcogenide glasses ($p_2 \sim 1$), confirming this region to be the IP.

The nearly stress-free nature of the IP is manifested in birefringence measurements using a polarizing microscope. Optically polished platelets of about 1mm thickness were examined with parallel polarizer and analyzer set up. Fig.4 reproduces micrographs of 6 sample compositions, two of which are below (10%,13%) , two in (16%,17%), and two above (20%,25%) the IP compositions. Black spots of about 50 μm in size are present in sample compositions outside the IP, but *not* in the IP. Furthermore, spot densities increase as one goes away from the IP compositions both below( 13% →10%) and above ( 20%→25%). Glasses are generally optically isotropic and can be expected to display no birefringence unless stress is frozen-in and renders a region optically anisotropic or birefringent. Our glass samples contain traces of *bonded* water as revealed by a band near 2.3 μm in mid-IR, and we believe the birefringent spots come from traces of *bonded* water that decorate macroscopic regions of frozen-in stress in the glasses. In samples synthesized by handling precursors in a glove box (rel. humidity less than 0.2%) bonded water content reduces substantially, and we were unable to observe birefringence in any of the elastic phases, suggesting that the intrinsic size of stressed regions is less than our resolution limit of 1 μm.

The present results suggest that additions of small amounts ( ~ 2%) of soda leads to creation of non-bridging oxygen atoms , and to the production of $Q^3$ species from $Q^4$ ones present in the base glass. Here $Q^n$ (NMR notation) represents a Ge having 'n' bridging oxygen nearest neighbors. The sharp drop in $T_g$ over the narrow interval, $0 < x < 2\%$, (Fig.1a) agrees with the slope, $dT_g/dx$ = - 30°C/at.% of soda, predicted by stochastic agglomeration theory[21]. With increasing soda content, five-fold coordinated Ge sites emerge[22] and increase the connectivity of the backbone as reflected in the increase of $T_g$ leading to a global maximum near x ~ 18%. We have modeled[23] the network structure by a speciation reaction[24] between a $Q^4$, a $Q^3$ and a five-fold Ge species ($Ge^{[V]}$) with respective probabilities $p_4$, $p_3$ and $p_5$. One is then able to define the mean Ge coordination number, $\boldsymbol{r_{Ge}} = 4p_4+3p_3+5p_5$ , and since $T_g$ correlates with network connectivity, we fix the reaction constant by requiring a maximum in $\boldsymbol{r_{Ge}}$ to coincide with that in $T_g$ (Fig.1a). The model then predicts (i) $T_g$s to decrease at x > 18% as $Q^3$ species proliferate and (ii) the mean-field rigidity transition to occur near x = 23%. Both predictions are in reasonable agreement with the data (fig.1). The broad global minimum in molar volumes near x ~18%, identified with the Germanate anomaly, has been attributed by some[13, 14, 22] to an aspect of local structure , i.e., Ge coordination number increase. On the other hand, Henderson[13] has suggested that the anomaly comes from growth of small $(GeO)_n$ rings, n = 3- and 4-member at the expense of larger ones, a feature of medium range structure. In Raman scattering one indeed observes the n = 3, 4 ring fraction to increase as x ~20%. The Germanate anomaly is better understood as a *direct consequence of the multiscale structural self-reorganization of the glasses leading to an IP* (Fig.1) in which local, intermediate and extended range structures are implied to fill available space compactly.

Another striking feature of our data are anomalies in IR response of LO ( $\nu_{LO}$(IR)) and TO ($\nu_{TO}$(IR)) mode frequencies shown in Fig. 3. Theoretical models of the IP suggest[4, 7] that its structure is dominated by the formation of non-crossing filamentary arrays with increasing ***r*** ( decreasing x). Here ***r*** = (8-6x)/3 if the bond-bending constraint centered about $Na^+$ ions is intrinsically broken. The rapid increase of $\nu_{LO}$(IR) with ***r*** in region I most likely reflects network stiffening due to increased cross-linking. The plateau of $\nu_{LO}$(IR) in region II reflects the (exponential) dominance of individual filamentary infrared coherence factors over linear filling factors. The jump in $\nu_{TO}$(IR) in region II (Fig.3b) reflects increased transverse stiffness of filamentary bundles achievable in the absence of fluctuating longitudinal internal network stresses. Overall the boundaries of the wide RW here in an oxide glass appear to be sharper than for similarly wide RW's in chalcogenide glasses, where sharp boundaries were observed only in the narrow RW of a (Ge,Se,I) alloy[25] with I network ends. A small concentration of Na and H network ends could facilitate both boundary sharpening and chain bundling.

In summary, we have provided evidence for existence of an IP in Sodium Germanate glasses, and have used birefringence measurements to confirm its nearly stress-free nature. The space filling property of the IP explains the global minimum in molar volumes and suggests an elastic origin for the Germanate anomaly. New IP related anomalies are observed in LO and TO vibrational modes, and these may assist in narrowing possible interpretations of that phase. This work is supported by NSF grant DMR 04- 56472. LPTMC is Unité Mixte de Recherche du CNRS n. 7600.

## Figure Captions

**Fig.1**. Compositional trends in (a) $T_g(x)$, (b) $\Delta H_{nr}(x)$ and (c) $\Delta C_p(x)$ from m-DSC. Also included in (a) are $T_g^{\eta}$ from ref. 18 , and the Ge mean coordination number, $\boldsymbol{r}_{Ge}$, from modeling and in (b) trends in $Q^3$, $Q^4$ and $Q^V$ species from modeling (c) molar volumes of glasses from density measurements on present samples.

**Fig.2**.IR based TO and LO response of glasses shown respectively in (a) $\varepsilon_2(\nu)$ and (b) of $-\mathrm{Im}(1/\varepsilon(\nu))$ at various glass compositions x indicated on the right in % .

**Fig.3**. (a) Changes in mode frequencies observed as functions of glass composition x. Note the mesa-like stiffening of $\nu_{TO}$(IR) in the IP.

**Fig.4**. Birefringence with an Olympus CX31 polarizing microscope. Black spots represent bonded water decorated regions of frozen-in stress in glass samples. Note the absence of spots for the two IP compositions (16% and 17%), but their growth with increasing density both below (13% and 10%) and above (20% and 25%) as one goes away from the IP.

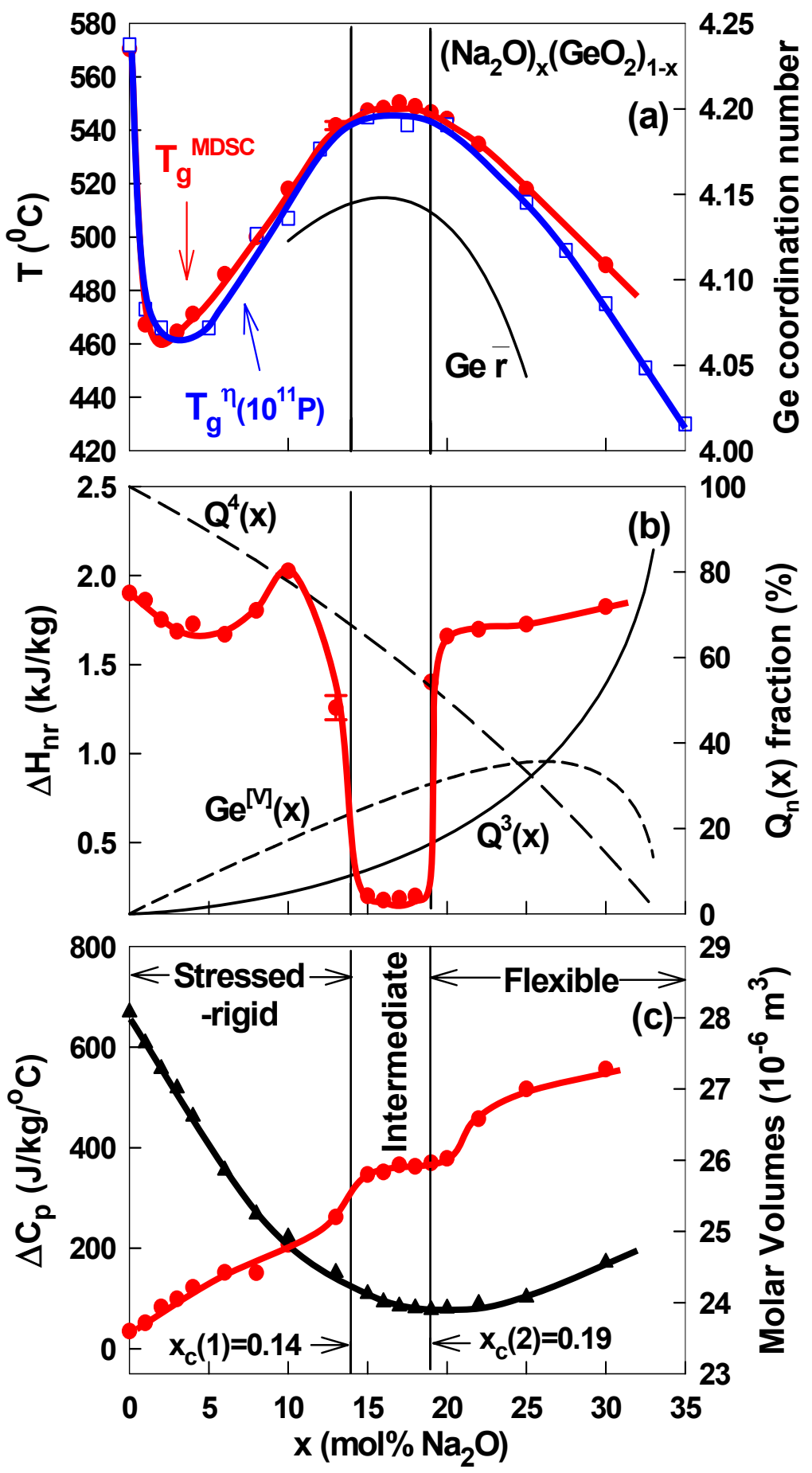

Figure 1

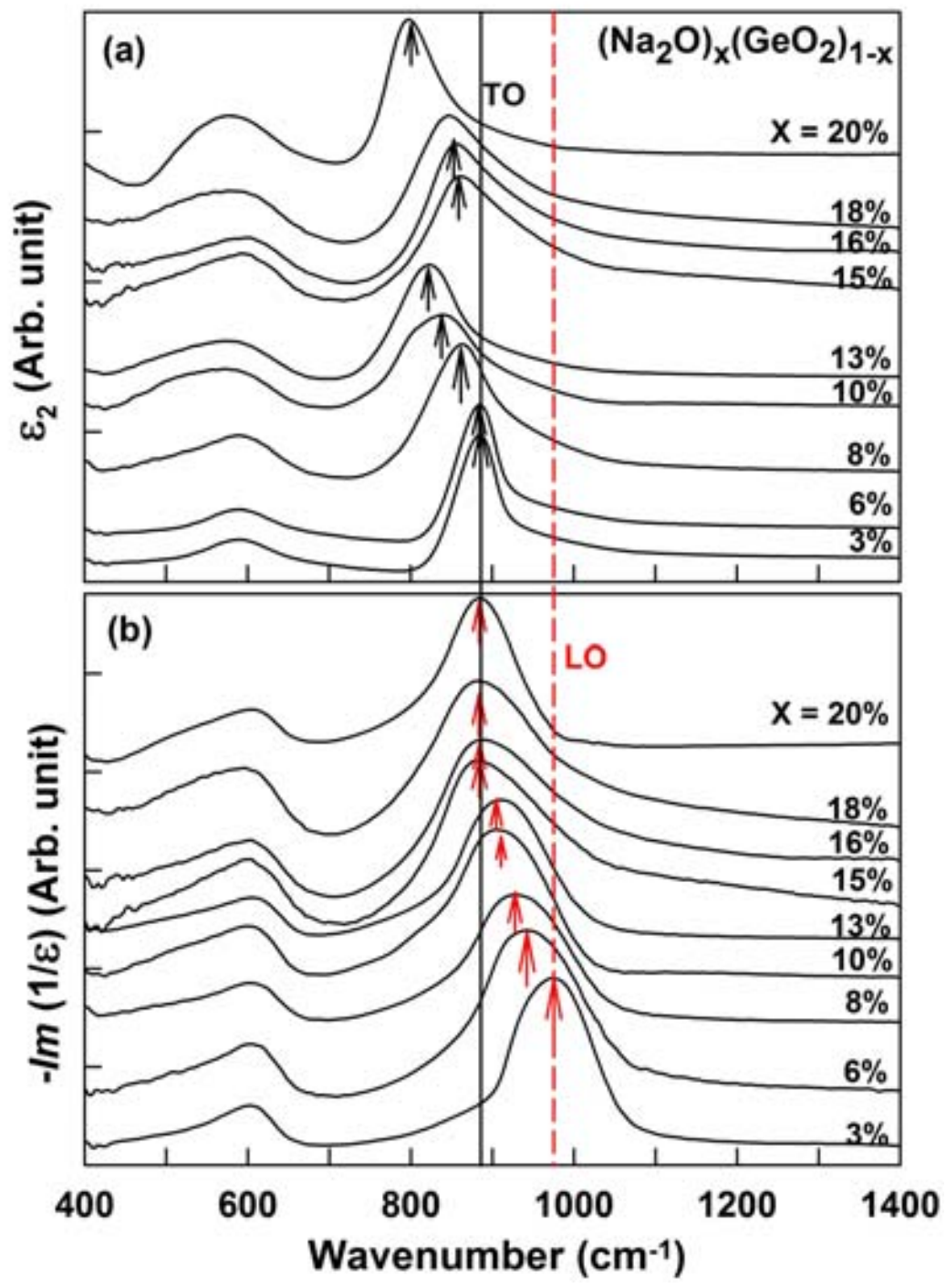

Figure 2

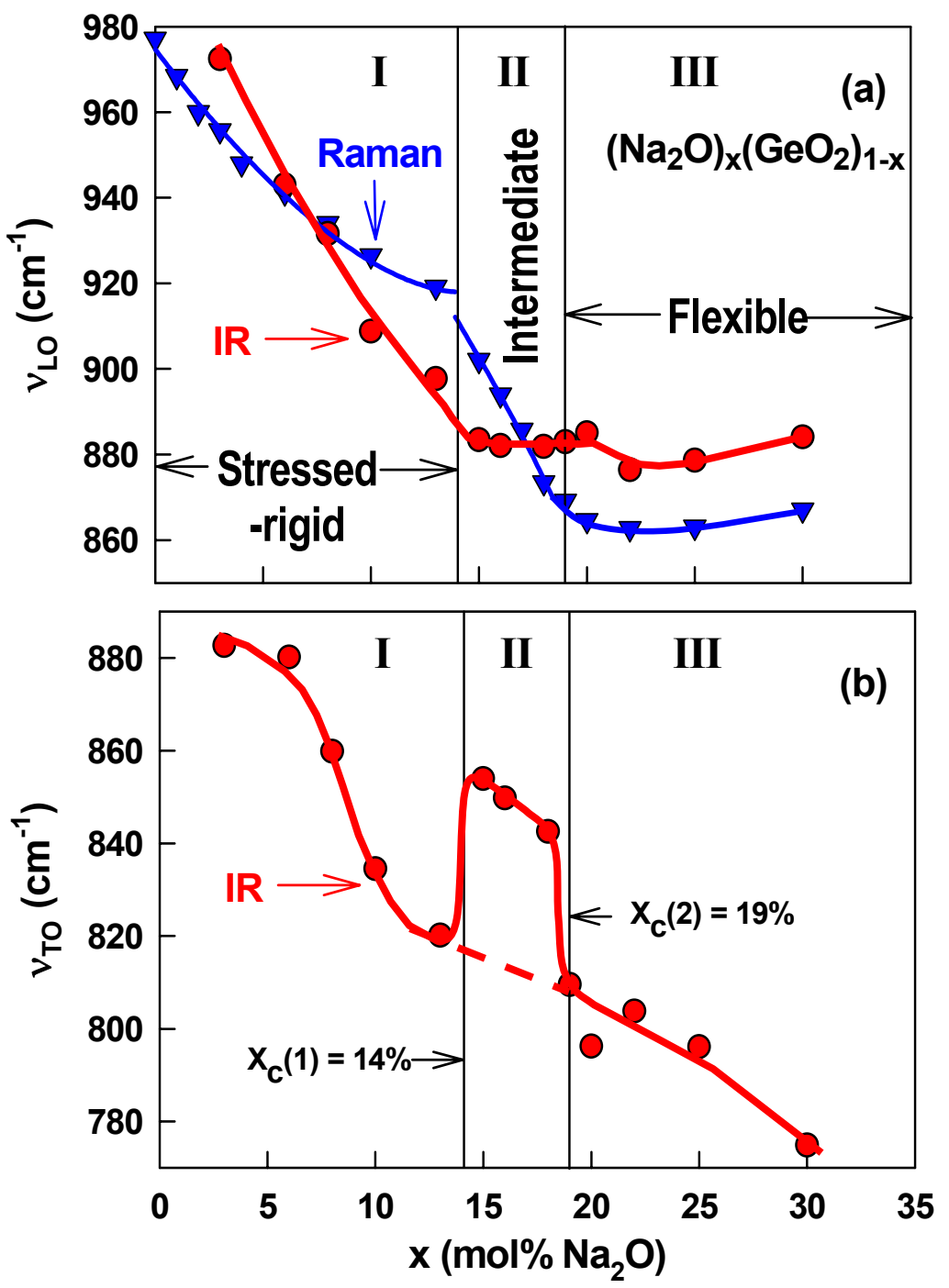

Figure 3

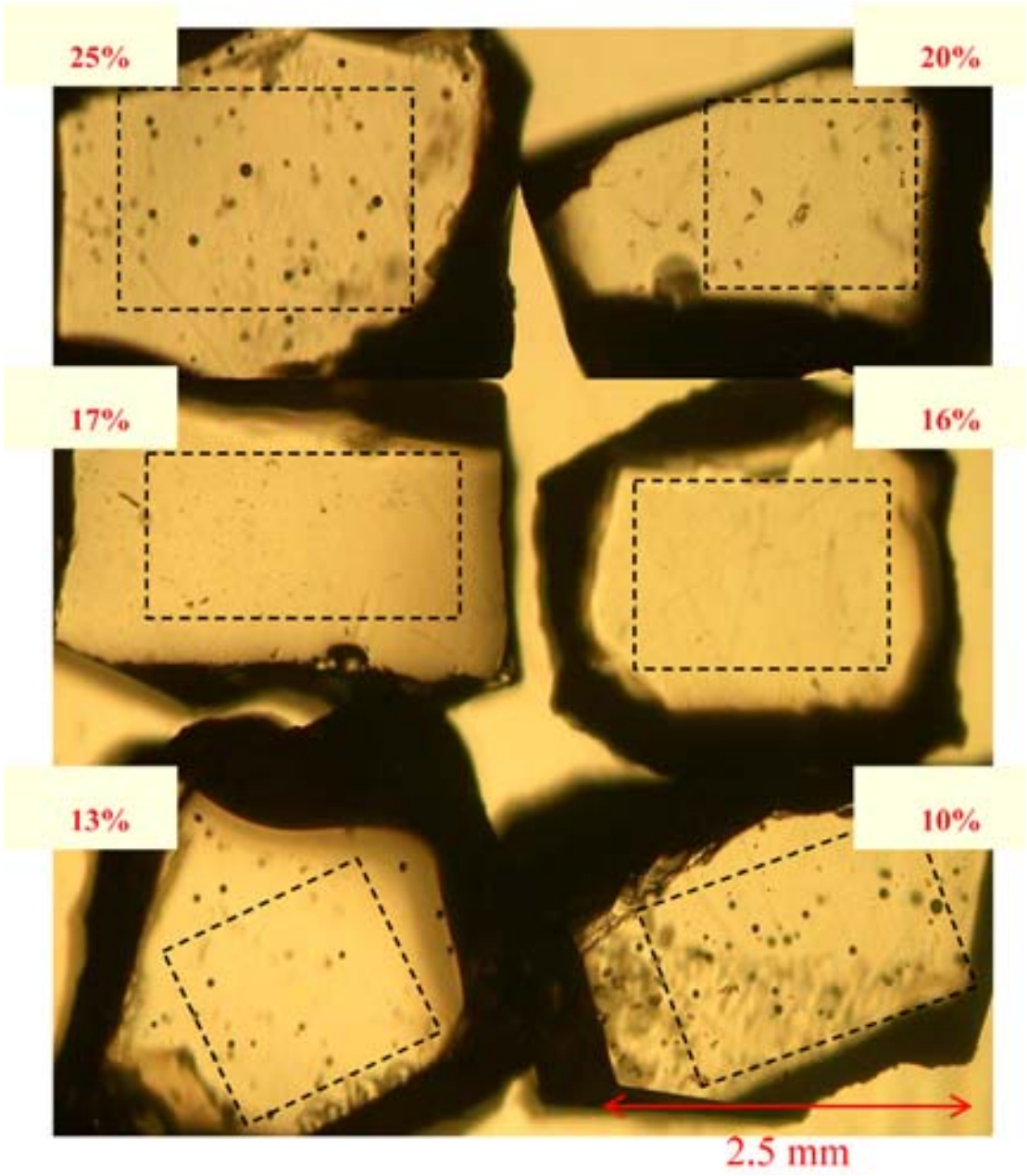

Figure 4